\documentclass[prl,twocolumn,showpacs,superscriptaddress]{revtex4}

\usepackage{graphicx}
\usepackage{amsfonts}
\usepackage{amsmath}
\usepackage{amssymb}

\begin{document}

\title{Nuclear signatures in high-harmonic generation from laser-driven muonic atoms}

\author{A.\ Shahbaz}
\affiliation{Max-Planck-Institut f\"{u}r Kernphysik, Saupfercheckweg 1, 69117 Heidelberg, Germany}
\author{C.\ M\"{u}ller}
\affiliation{Max-Planck-Institut f\"{u}r Kernphysik, Saupfercheckweg 1, 69117 Heidelberg, Germany}
\author{A.\ Staudt}
\affiliation{Max-Planck-Institut f\"{u}r Kernphysik, Saupfercheckweg 1, 69117 Heidelberg, Germany}
\author{T.\ J.\ B\"{u}rvenich}
\affiliation{Frankfurt Institute for Advanced Studies, Johann Wolfgang Goethe University, Max-von-Laue-Str. 1, 60438 Frankfurt am Main, Germany}
\author{C.\ H.\ Keitel}
\affiliation{Max-Planck-Institut f\"{u}r Kernphysik, Saupfercheckweg 1, 69117 Heidelberg, Germany}

\begin{abstract}
High-harmonic generation from muonic atoms exposed to intense laser fields is considered. Our particular interest lies in effects arising from the finite nuclear mass and size. We numerically perform a fully quantum mechanical treatment of the muon-nucleus dynamics by employing  modified soft-core and hard-core potentials. It is shown that the position of the high-energy cutoff of the harmonic spectrum depends on the nuclear mass, while the height of the spectral plateau is sensitive to the nuclear radius. We also demonstrate that $\gamma$-ray harmonics can be generated from muonic atoms in ultrastrong VUV fields, which have potential to induce photo-nuclear reactions.
\end{abstract}
\maketitle

%
Muonic atoms have ever since been useful tools for nuclear spectroscopy employing atomic-physics techniques. Due to the large muon mass and the correspondingly small Bohr radius, the muonic wave function has a large overlap with the nucleus and thus effectively probes its structural features, such as finite size, deformation, surface thickness, or polarization \cite{EiGr1970}. Muonic atoms also play an important role as catalysts for nuclear fusion \cite{BrKaCoLe1989}.
On the other hand, there is the steadily growing field of laser-matter interactions. While traditionally dealing with the atomic or molecular response to an external light wave, it has recently widened to include also, among other new directions, laser-nuclear physics \cite{ScMaBe2006}. With laser-heated clusters and plasmas, nuclear fission \cite{LeCo2000} and fusion \cite{Di1997Di1999} have been realized as well as neutron production in (p,n) or (d,n) reactions \cite{Um2003LeMcSi2003}. By virtue of the interaction with an applied laser field, high-precision experiments have been able to monitor ultrafast vibrations of the nuclei in diatomic molecules \cite{Ba2006}. The rapid progress in laser technology even opens prospects for nuclear quantum optics via direct laser-nucleus coupling \cite{1BuEvKe2006}.

Against this background, the combination of muonic atoms with laser fields in order to \textit{dynamically} probe nuclear properties by the laser-driven muon appears very promising. The information on the nucleus, dynamically revealed by laser assistance, can in principle complement the knowledge gained from field-free spectroscopy of muonic atoms. Modifications of muon-catalyzed fusion from a muonic $D^{+}_{2}$ molecule in the presence of a superintense laser field have recently been addressed \cite{ChBaCo2004}. We note that on the typical time scale of strong laser pulses ($\tau \sim$ fs$-$ns) muonic atoms and molecules may  be considered as stable since the muon lifetime is 2.2 $\mu$s.

%
In this Letter, high-harmonic generation (HHG) \cite{BeMi2002} from muonic hydrogen isotopes in strong laser fields is studied. We demonstrate nuclear signatures arising from the finite nuclear mass and size in the radiation emitted by the muonic charge cloud, which oscillates across the nucleus under the influence of the laser field. The relative motion of the muon and the nucleus is taken into account. In the ground state of muonic hydrogen, the muon is bound by 2.5 keV and experiences an electric field strength of $1.8\times10^{14}$ V/cm corresponding to the intensity $4.2\times10^{25}$ W/cm$^{2}$. It is of interest to compare these numbers with the parameters of the most advanced present-day and near-future laser sources. In the optical and near-infrared frequency range ($\hslash \omega \sim 1$ eV), the next generation of high-power lasers will reach the intensity level of $10^{23}$ W/cm$^{2}$ \cite{MoTaBu2006}. At VUV frequencies ($\hslash \omega \sim 10-100$ eV) a record intensity of almost $10^{16}$ W/cm$^{2}$ has recently been achieved at the FLASH facility (DESY, Germany) with a free-electron laser \cite{web2006}. Near-future upgrades of such machines are envisaged to generate x-ray beams ($\hslash \omega \sim 1-10$ keV) with peak intensities close to $10^{20}$ W/cm$^{2}$. There are also efforts to produce ultrashort, high-frequency radiation ($\hslash \omega \sim 10-1000$ eV) from plasma surface harmonics where, due to high conversion efficiency, considerably higher intensities might be reachable \cite{Ts2006}. With these high-intensity and/or high-frequency sources of coherent radiation it will become possible to influence the dynamics of (light) muonic atoms. As an example, we investigate high-harmonic emission from muonic hydrogen and deuterium in intense VUV laser fields. Our numerical calculations reveal pronounced differences in the HHG spectrum when changing the nuclear mass and size.

%
We consider muonic quantum dynamics in the non-relativistic regime described by the time-dependent Schr\"{o}dinger equation (TDSE). Due to the large muon mass, the nucleus cannot be treated as an infinitely heavy particle and its motion must be taken into account. Within the dipole approximation for the laser field, the two-particle TDSE separates into center-of-mass and relative motion. By employing a one-dimensional (1D) hydrogen-like model atom (to be justified below), the center of mass evolves in time according to
\begin{equation}
\label{CMSE}
i\hslash\dfrac{\partial}{\partial t}\Psi\left( X,t\right)=\left[\dfrac{P^{2}}{2M}+(Z-1)eXE\left(t\right)\right]\Psi\left( X,t\right)
\end{equation}
and the relative motion is governed by
\begin{widetext}
\begin{equation}
\label{RSE}
i\hslash\dfrac{\partial}{\partial t}\psi \left(x,t\right)=\left[\dfrac{p^{2}}{2m_{r}}-m_{r}\left(\dfrac{Z}{m_{n}}+\dfrac{1}{m_{\mu}}\right)exE(t)+V\left(x\right)\right]\psi \left(x,t\right)%
\end{equation}
\end{widetext}
with the muonic, nuclear, total and reduced masses $m_{\mu}$, $m_{n}$, $M=m_{\mu}+ m_{n}$ and $m_{r}=m_{\mu} m_{n}/M$, respectively, center-of-mass and relative coordinates $X$ and $x$, corresponding momentum operators $P=-i\hslash\partial/\partial X$ and $p=-i\hslash\partial/\partial x$, nuclear charge number $Z$, nuclear potential $V(x)$ and laser electric field $E(t)$. Equation (\ref{CMSE}) is the non-relativistic Volkov equation for a particle of charge $(Z-1)e$ and mass $M$. In the present paper we restrict ourselves to the case $Z=1$ as the required laser intensities are the smallest then, while the effects of interest are expected to be appreciable. Note that throughout the nuclear chart the largest relative difference among isotopes, both in mass and charge radius, exists between hydrogen nuclei. In this situation, Eq.\,(1) describes free motion of the center-of-mass coordinate which does not generate radiation and can be ignored. The laser field solely couples to the relative coordinate. For $Z=1$, Eq.\,(\ref{RSE}) is reduced to the usual Schr\"{o}dinger equation for a single particle of charge $-e$ and mass $m_{r}$ in the combined fields of a laser and a nucleus. By applying the standard scaling transformation, one can recast Eq.\,(\ref{RSE}) into a form that describes an ordinary (\textit{i.e.} electronic) hydrogen atom in a scaled laser field. The scaled laser frequency and field strength are given by \cite{ChBaCo2004}
\begin{eqnarray}
\label{scaling}
\omega^\prime=\rho\omega\hspace{1.2cm};\hspace{0.5cm}E^\prime=\rho^{2}E
\end{eqnarray}
where $\rho \equiv m_{e}/m_{r}$, with the electron mass $m_{e}$. That is, a muonic hydrogen atom in a laser field with parameters $E,\omega$ behaves like an ordinary hydrogen atom in a field with $E^\prime,\omega^\prime$ given by Eq.\,(\ref{scaling}), provided that the Coulomb potential $V(x)$ arises from a pointlike nucleus. The scaling procedure, however, does not account for nuclear parameters like, \textit{e.g.}, the finite nuclear size. Evidently, when the transition from a muonic to an ordinary hydrogen atom is performed, the nuclear radius is not to be scaled but remains fixed. Hence, for atomic systems where nuclear properties play a role, not all physical information can be obtained via scaling. Below we show results which display the influence of the finite nuclear size on the process of HHG.

We solve the TDSE in Eq.\,(\ref{RSE}) numerically by making use of the Crank-Nicolson time-propagation scheme. The laser field is always chosen as a 5-cycle pulse of trapezoidal envelope comprising one cycle for linear turn-on and turn-off each. The harmonic spectrum is obtained by taking the Fourier transform of the dipole acceleration. With regard to the 1D approximation in Eqs.\,(\ref{CMSE},\ref{RSE}) we note that, in general, 1D models are known to retain the essential physical features of nonrelativistic laser-atom interaction for linearly polarized fields \cite{EbBook1992}; therefore, these models are widely used \cite{FaKoBeRo2002}. Furthermore, in our case the 1D numerics still are a non-trivial task because of the fine grid spacing required to resolve the nuclear extension and the non-standard laser parameters employed. As regards ordinary atoms, the latter would correspond to intense fields in the far-infrared. Eventually, we stress that the goal of this paper is to reveal \textit{relative} differences between physical observables which typically are less sensitive to model assumptions than absolute numbers. The main shortcoming of the 1D approach consists in neglecting the muon's transversal wave-packet spreading which can substantially reduce the total harmonic yield \cite{spreading}.

%
As the conversion efficiency of HHG is rather low, it generally is desirable to maximize the harmonic signal strength. In the present case, the optimization is of particular importance as the target density is low. Efficient HHG requires both efficient ionization and recombination. The former is guaranteed if the laser peak field strength lies just below the border of over-barrier ionization (OBI) where the Coulomb barrier is suppressed all the way to the bound energy level by the laser field, \textit{i.e.}
\begin{equation}
\label{OBI}
E\lesssim E^{\rm OBI} = \dfrac{m_{r}^{2}c^{3}}{e\hbar}\cdot\dfrac{\alpha^{3}}{16}\,,
\end{equation}
corresponding to the intensity $I^{\rm OBI}\approx1.6\times10^{23}$ W/cm$^{2}$. Here, $\alpha \approx 1/137$ denotes the fine-structure constant. Efficient recombination is ensured if the magnetic drift along the laser propagation direction is negligibly small, which limits the relativistic parameter to \cite{WaKeScBr2000}
\begin{equation}
\label{RP}
\xi\equiv\dfrac{eE}{m_{r}c\omega}<\left(\dfrac{16\hbar \omega}{\sqrt{2m_{r}c^{2}I_{p}}} \right)^{\frac{1}{3}},
\end{equation}
with the atomic ionization potential $I_{p}$. We note that condition (\ref{RP}) also implies applicability of the dipole approximation. The lowest frequency which simultaneously satisfies Eqs.\,(\ref{OBI}) and (\ref{RP}) lies in the VUV range and corresponds to $\hbar\omega\approx27$ eV. At this value, the harmonic radiation has a remarkable maximum energy of $E_{c}\approx0.55$ MeV. Hence, muonic atoms are also promising candidates for the generation of hard x-rays which might be employed to trigger photo-nuclear reactions.

%

\begin{figure}
\includegraphics*[width=0.64\linewidth,angle=270]{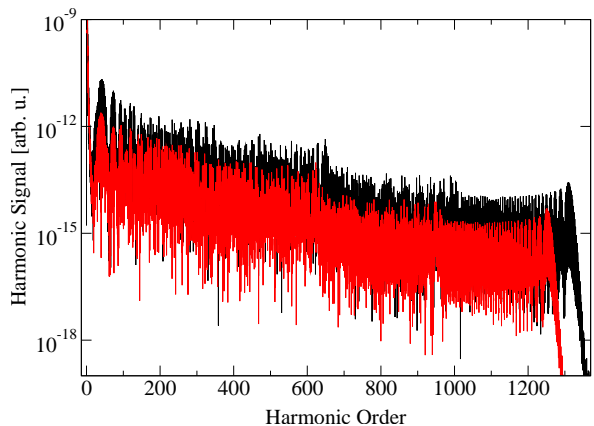}
\caption{(color online). HHG spectra calculated with the soft-core potential (\ref{SC}). The black and grey (red) lines represent the spectrum for muonic hydrogen and deuterium, respectively. The laser parameters are $I=1.05\times10^{23}$ W/cm$^{2}$ and $\hslash \omega=59$ eV.}
\label{fig1}
\end{figure}
\begin{figure}[<h>]
\includegraphics*[width=0.64\linewidth,angle=0]{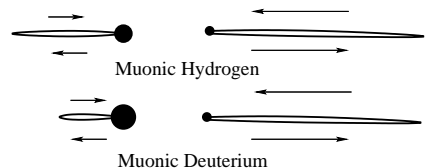}
\caption{Schematic diagram of the motion of the nucleus (on the left) and the muon (on the right) in muonic hydrogen and deuterium in the presence of a laser field polarized in horizontal direction. The oscillation amplitudes are not scaled.}
\label{fig2}
\end{figure}

To express the possible effects due to the finite nuclear mass appearing in HHG spectra of muonic hydrogen isotopes, we introduce the usual soft-core potential \cite{EbBook1992}
\begin{equation}
\label{SC}
V_{s}(x)=-\dfrac{1}{\sqrt{x^{2}+\rho^{2}}}\,,
\end{equation}
appropriately scaled to the muonic system. Here, we consider only the variation of nuclear mass assuming the nucleus as being point-like. Figure \ref{fig1} shows the HHG spectra for muonic hydrogen and deuterium in an ultra-strong VUV field of the frequency $\hslash \omega=59$ eV and the intensity $I=1.05\times10^{23}$ W/cm$^{2}$. The spectrum of hydrogen is enhanced and further extends by about 60 harmonics. The reason for the spectral enhancement is that the effective particle of the proton-muon system experiences stronger acceleration since it has a smaller reduced mass. The difference in the cutoff position $E_{c}=I_{p}+3.17U_{p}$ \cite{BeMi2002} is also caused by the reduced mass which enters the ponderomotive energy
\begin{equation}
\label{UP}
U_{p}=\dfrac{e^{2}E^{2}}{4\omega^{2}m_{r}}=\dfrac{e^{2}E^{2}}{4\omega^{2}}\left(\dfrac{1}{m_{\mu}}+\dfrac{1}{m_{n}}\right).
\end{equation}
For the chosen laser parameters we obtain $U_{p}\gg I_{p}$ so that the smaller reduced mass leads to a higher cutoff energy. Note that in a situation where $U_{p}\ll I_{p}$, the relative order of the cutoff positions would be reversed. Furthermore, we point out that the pondermotive energy in Eq.\,(\ref{UP}) can be understood as the sum of the ponderomotive energies of muon and nucleus. As schematically depicted in Fig.\,\ref{fig2}, both particles are driven into opposite directions by the laser field and when they recollide their kinetic energies add up. In this picture the higher cutoff energy of the hydrogen atom arises from the larger ponderomotive energy of the proton as compared to the deuteron. Note that the phenomenon that both atomic constituents are driven into opposite directions is most pronounced for positronium, the bound state of an electron and positron \cite{HeHaKe2004}. We also observe a few kinks in the spectra of Fig.\,\ref{fig1}. This multiplateau structure is typical of the interaction regime where $\xi\sim0.1$ (in our case $\xi\approx0.04$) \cite{BeMi2002}.

At the chosen frequency, the laser intensity assumed in Fig.\,\ref{fig1} cannot be attained by present-day technology in the laboratory. One might, however, employ a relativistic beam of muonic atoms counter-propagating the laser beam \cite{ChJoKyPo2004}. Then, via the relativistic Doppler shift, the parameters of Fig.\,\ref{fig1} can be reached by means of an intense optical laser pulse of $\hbar\omega\approx1.5$ eV and $I\approx6.5\times10^{19}$ W/cm$^{2}$ at an atomic Lorentz factor of $\gamma\approx20$. In their rest frame, the atoms would then experience the laser parameters of Fig.\,\ref{fig1}. For muonic hydrogen atoms carrying zero charge, the acceleration cannot directly be accomplished, though. But one could accelerate muons and protons/deuterons separately and let them combine in a merged beams setup. A similar way of producing positronium atoms at $\gamma\approx20$ via acceleration of negative positronium ions has been proposed recently \cite{Ug2006}.

\begin{figure}
\includegraphics*[width=0.64\linewidth,angle=270]{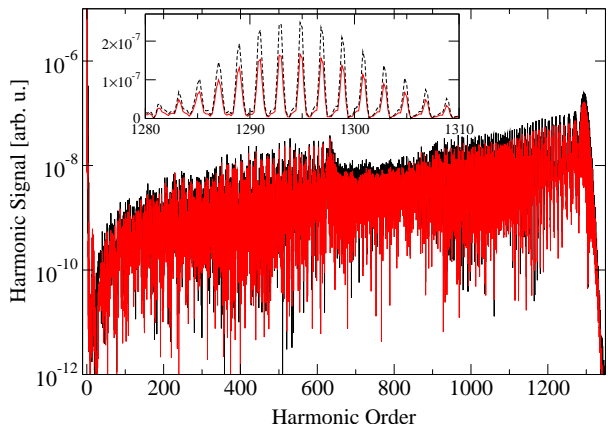}
\caption{(color online). HHG spectra calculated with the hard-core potential (\ref{HC}). The black line represents the spectrum for muonic hydrogen at the laser parameters $I^{\rm H}=1.05\times10^{23}$ W/cm$^{2}$ and $\hslash \omega^{\rm H}=59$ eV. The grey (red) line represents the spectrum for muonic deuterium at the correspondingly scaled laser parameters $I^{\rm D}=1.30\times10^{23}$ W/cm$^{2}$ and $\hslash \omega^{\rm D}=62$ eV. The inset shows a blow-up of the cutoff region on a linear scale, with the black line dashed for better visibility.}
\label{fig3}
\end{figure}

%
To express the possible effects due to the finite nuclear size appearing in HHG spectra of laser-driven muonic hydrogen isotopes, we need a potential which explicitly takes the nuclear radius into account. We make use of the nuclear drop model, considering the nucleus as a sphere of uniform charge density within the nuclear radius $R$, and employ the potential:
\begin{equation}
\label{HC}
V_{h}(x)=\begin{cases} -\dfrac{1}{R}\left( \dfrac{3}{2}-\dfrac{x^{2}}{2R^{2}}\right)  &  \text{if $|x|\leq R$,}
\\
-\dfrac{1}{|x|} &\text{if $|x|>R$.}
\end{cases}
\end{equation}
We note that the lowest lying atomic state of this potential is unphysical as its binding energy becomes infinite in the limit $R\rightarrow0$ \cite{ScFa1994}. Therefore, in accordance with common practice \cite{ScFa1994,Ch1996,GoSaKa2005}, we start our calculation from the first excited state which has the correct hydrogenic binding energy. Figure \ref{fig3} shows the HHG spectra obtained from the potential (\ref{HC}) for muonic hydrogen and deuterium. The proton and deuteron charge radii are $R_p\approx 0.875$ fm and $R_d\approx 2.139$ fm \cite{MoTa2005}. In order to compensate for the nuclear mass effect, we apply accordingly scaled frequencies and intensities with $\omega \propto m_{r}$ and $I\propto m_{r}^{4}$ [cf. Eq.\,(\ref{scaling})] so that the cutoff positions of both spectra coincide. We note that the overall shape of the harmonic response differs from that shown in Fig.\,\ref{fig1} due to the different potentials employed. In particular, a dip can be seen at low harmonics ($n \hslash\omega\sim I_{p}$) followed by a rising plateau region. A similar feature was found in \cite{HuStBeSaMi2002} and attributed to the behavior of non-tunneling harmonics in very steep potentials. More important with regard to the present study is, however, that the harmonic signal from muonic hydrogen is larger (by about 50 \% in the cutoff region) than that from deuterium. The reason is that a smaller nuclear radius increases the steepness of the potential near the origin, leading to more violent acceleration and thus to enhanced harmonic emission. With regard to ordinary atoms it has been noticed before that the analytic behavior of the binding potential near the origin is of great importance for the HHG process \cite{GoSaKa2005}; according to Ehrenfest's theorem, the atomic dipole is accelerated by the potential gradient which is largest in this region. In the present situation we therefore find that the plateau height of HHG spectra from muonic atoms is sensitive to the finite nuclear size.

%
In conclusion, we have found characteristic nuclear signatures in the harmonic response of strongly laser-driven muonic hydrogen and deuterium atoms. For deuterium, due to the larger nuclear mass and size, both the harmonic cutoff and the plateau height are significantly reduced. It was also shown that high-energy photons in the MeV regime can be generated which are capable of nuclear excitation. Our results could experimentally be tested by virtue of near-future technology, \textit{i.e.} by employing intense high-frequency radiation from free-electron lasers or plasma surface harmonics or, alternatively, by combining radiation from existing high-power near-optical laser systems with a relativistic atomic beam. The present study demonstrates that muonic atoms in strong laser fields can be utilized to dynamically gain structure information on nuclear ground states. It also offers the prospect of performing pump-probe experiments on excited nuclear levels since the periodically driven muon can excite the nucleus during one of the encounters and afterwards probe the excited state and its deexcitation mechanism.

%
A. Shahbaz acknowledges support by the Higher Education Commission (HEC), Pakistan and Deutscher Akademischer Austauschdienst (DAAD).

%

\end{document}